\documentclass[%
 reprint,
 amsmath,amssymb,aps,
]{revtex4-1}

\usepackage{graphicx}
\usepackage{dcolumn}
\usepackage{bm}

\begin{document}


\title{Gamma rays from dark mediators in white dwarfs}

\author{M. Cerme\~no}
 \email{marinacgavilan@usal.es}
\author{M. A. P\'erez-Garc\'ia}%
 \email{mperezga@usal.es}
\affiliation{%
 Department of Fundamental Physics, University of Salamanca, Plaza de la Merced s/n, E-37008 Salamanca, Spain
}%


\date{\today}

\begin{abstract}
We consider self-annihilation of dark matter, $\chi$, into metastable mediators, $Y$, and their subsequent decay into photons inside white dwarfs. We focus on reactions of the type $\chi \bar{\chi}\rightarrow YY$, where mediators, besides having a finite decay lifetime at rest $\tau_{\rm rest}\lesssim 1$ s, may suffer energy loss in the medium before they decay into photons, $Y \rightarrow \gamma\gamma$. We obtain attenuated gamma-ray luminosities arising from the combination of both effects. Using  complementary sets of astrophysical measurements from cold white dwarfs in the M4 globular cluster as well as direct/indirect dark matter searches we  discuss further constraints on dark mediator lifetimes.
\end{abstract}

\maketitle


\section{Introduction}
Dark matter (DM) accumulation sites can provide a valid strategy to potentially identify hints of its existence as well as about its nature and properties. In particular, one could think of concentration of this type of matter inside astrophysical bodies as a consequence of gravitational interaction and, provided suitable ranges of masses and cross sections, thermalization with ordinary --baryonic-- matter constituting these objects. Effects such as self-annihilation, co-annihilation with a different species or decay have been exhaustively studied as possible multi-messenger signatures taking place in the Sun, planets, main sequence stars or more compact objects like white dwarfs or even neutron stars \cite{silkpho, spergel, Ritz, kouv2, gould,fairbairn, perez1,cer1,cer2}.

Regarding DM itself, Beyond Standard Model (BSM) candidates have flourished in the literature over the past decades, for a review see, for example \cite{bertone}. Some of them are now well constrained from the coordinated effort of different communities focusing on direct, indirect and collider searches \cite{summer}. One of the currently accepted possible realizations considers that DM could generate a relic density via annihilation into so-called {\it dark mediators} and subsequent SM particles.
These type of models are also referred to as {\it secluded} in the sense that DM dramatically reduces its couplings to SM states by an intermediate state, a decaying mediator, $Y$. Examples in the literature include coupling strengths ranging from weakly interacting DM particles \cite{pospe} to  strongly interacting ones \cite{hochberg}. According to the duration of lifetime for these metastable particles, there is a further division among short-lived or long-lived mediators, each giving rise to dramatic differences in the predicted indirect signal \cite{roth, lucente}.
Annihilation of DM into two generic different mediators $\chi {\bar \chi} \rightarrow M_1 M_2$ can take place in the s- or p-wave scheme, depending on the Lorentz structure of the DM-mediator interactions \cite{bell2}. There are some works that have focused on particular realizations of these secluded models \cite{plehn}. 

Indirect signals from DM could also be expected from the possibility of annihilation through long-lived mediators into gamma rays in astrophysical environments, the galactic center, dwarf spheroidals and the CMB \cite{pospe,escudero,profumo,Chu} and also into neutrinos \cite{bell-petra,ardid}.  
From the experimental side, current DM searches are actively constraining the available mass and DM-nucleon cross-section,  $m_\chi-\sigma_{\chi,N}$, phase space. Efforts include those from colliders like e.g. Belle \cite{belle} or LHC \cite{plehn}. Regarding indirect searches \cite{cirelli} in the neutrino channel secluded DM models have been constrained by ANTARES \cite{antares} while in the gamma-rays relevant searches are performed by Fermi, H.E.S.S. and AMS collaborations \cite{satellite}.

On general grounds, considering dark metastable mediators enriches the picture by which DM interacts with ordinary SM matter. The main characteristic in this scenario comes from the fact that these particles  have a finite lifetime. While short-lived mediators would be essentially indistinguishable in most of the searches from models where DM is not secluded, long-lived mediators allow an injection of SM states not directly related to places with enhanced DM density.
Therefore, it could produce detectable signals far from the production site. Besides, this mechanism has been quoted \cite{Chu} to introduce associated anisotropies of prompt species -- positrons and photons-- produced in the decays of long-lived mediators. Popular mediators such as the dark photon or the dark Higgs have been featured in a number of recently proposed Dark Sector Models \cite{arkani} although several model independent DM scenarios also feature long-lived particles \cite{tha}.

In this work we consider the annihilation of light DM particles inside white dwarfs (WDs). We restrict our interest to dark candidates in the sub-GeV mass range as they can probe some particularly interesting  astrophysical scenarios \cite{cline}. Recent constraints show a window for masses $m_\chi\lesssim 1$ GeV and $\chi-N$ cross sections $\sigma_{\chi,N} \lesssim 10^{-29}$ $\rm cm^2$ (see Fig.6 in \cite{kavana}). Electrons are also relevant species inside WDs. However, their scattering cross-section with sub-GeV DM is typically much smaller than that of nucleons \cite{elec} (less than $\sim 10^{-38}$ $\rm cm^2$ for thermal DM where form factors do not play a role). Let us emphasize at this point that since the mass range  beyond a few GeV is nowadays better probed, the less constrained light mass phase space seems the next interesting region to explore \cite{hooper} with proposals allowing strongly interacting candidates (SIMPs) as recently described in \cite{collar}.

The astrophysical scenario we consider is that of a WD where DM particle annihilates inside the stellar medium so that the metastable mediators produced in the reaction $\chi \bar{\chi}\rightarrow YY$, can loose energy while propagating (on-shell) outwards. Eventually they will decay into photons, $Y \rightarrow \gamma\gamma$, either inside or outside the star as this is governed by their energy-dependent lifetime and dissipation.
White dwarfs are compact stars made mostly of Carbon and Oxygen and formed at the end of the lifetime of main sequence stars with masses up to $\sim8 M_\odot$. They represent around $\sim95\%$ of all the stars in our galaxy. It is believed that a fraction up to $20\%$ may harbour magnetic fields with a strength up to several hundred MG. Since at their final stages no fusion reactions can happen in their interior they are supported by electron pressure. Typical masses range $M_{WD}\sim (0.3-1) M_\odot$ and radii $R_{WD}\sim (0.01-0.03)R_\odot$. Thus, these stars are essentially cold, with effective temperatures $T\sim 10^3-10^4$ K, and dense enough $\rho \sim 10^{4\div7}$ $\rm g/cm^3$ so that density effects can not be neglected.
In order to avoid much of the limitations set by particular model settings we will consider a generic scenario of annihilating fermionic DM into photons via dark mediators making a minimal set of assumptions. Depending on how long or short-lived these mediators are and the importance of dissipation we will show how gamma ray emission and luminosities from WDs can be attenuated. For example, if the mediator lifetime is large enough, it could decay outside the stellar radius modifying the expected energy flux value with respect to that arising from decay at central regions. 
In this scenario, medium effects have to be dealt with as, generically, a mediator will loose energy passing through the ordinary matter provided the decay length, $\lambda_D$, is larger than the interaction length, $\lambda_I$. These boosted mediators will be produced with initial velocity, $v_{Y,0}$, that will be attenuated while propagating inside the star thus affecting its survival probability and the energy deposited at the decaying site. However, let us emphasize that for less dense stellar bodies such as the sun or planets, average densities are less than $\sim 10^2$ $\rm g/cm^3$, thus much lower than central densities in WDs and dissipation effects are less important.

The paper is structured as follows. In section II we describe the DM annihilation into dark mediators as well as particle energy losses inside the WD scenario. We calculate the survival probabilities for mediators with energy dependent lifetimes, their spectrum and the associated photon luminosities. In section III we  show results concerning gamma-ray luminosities comparing to those for cold WDs in M4 globular cluster (GC) and further discuss possible constraints on dark mediator lifetimes. Finally, in IV we conclude.

\section{Dark mediators and in-medium interaction}

In this section we describe the process where the photon emission arises from DM annihilation via dark mediators. We consider fermionic dark matter particles, $\chi$, that annihilate into metastable mediators, $Y$, through reactions $\chi \chi \rightarrow YY$ with subsequent decay $Y \rightarrow 2\gamma $. Additional reactions from radiative emission processes or three body decay of mediators \cite{Kim} induce small corrections,  including anisotropies \cite{Chu} that we will not consider here. 
It is well known that considering DM candidates not restricted to weakly interacting particles introduces the possibility that they could undergo self-interactions with 3-to-2 or 4-to-2 annihilations \cite{hochberg,herms}, however we expect our results will not be  qualitatively modified as stellar DM densities remain small. 

The metastable mediator has a lifetime at rest that can be related to its decay width as $\tau_{\rm rest}=\Gamma^{-1}_Y$ (we use  $\hbar=c=1$). As  initially created in the boosted $2\rightarrow 2$ process, they possess a Lorentz factor $\gamma_{Y,0}=\frac{1}{\sqrt{1-v_{Y,0}^2}}=m_\chi/m_Y$. In the stellar environment the mediator will, in principle, interact with the medium decreasing its energy and velocity, $v_Y$, from initial values. Attenuation of dark candidates has been considered as a source of DM depletion on terrestrial detectors \cite{starkman,kouv3}. This stopping power is a crucial aspect that could largely impact the energetic yields of annihilation processes in a medium.  

From the accumulated DM present inside the WD, dark mediators are produced from the two body annihilation reaction $\chi \bar{\chi}\rightarrow YY$. Regarding the kinematics of the dark mediator propagation we will use a  radial treatment inside the star for the light DM mass range and (relatively) strong interacting cross sections motivated by  previous findings of small-deflection angle approximation \cite{starkman}. In the medium they may suffer from interaction with ordinary matter composed of nuclei and a gas of electrons. Such a possibility is realized when they live long enough to experience the scattering processes we consider, e.g. if their decay length  $\lambda_D$ is larger than the interaction length, $\lambda_{I}\sim \frac{1}{\sigma_{Y,i} n_i}$. $n_i$ is the number density of scattering centers of $i$th-type in the stellar volume (nuclei of baryonic number $A$ end electrons) and $\sigma_{Y,i}$ is the scattering cross-section describing mediator interactions with those matter constituents. Besides, the possible interaction with electrons $\sigma_{Y,e}$ is unknown but we take it to be bounded by that of DM, i.e. $\sigma_{Y,e}\lesssim \sigma_{\chi,e}$ according to experimental constraints \cite{elec}. In the situation where scattering of DM with electrons is less frequent than for nuclei $\lambda_{\chi,e}>\lambda_{\chi,A}$ or, equivalently, $\sigma_{\chi, e}\lesssim 2 A \sigma_{\chi,N}$, the effect of electrons can be safely neglected in the sub-GeV mass range. In this work we will constrain all decaying mediators to have rest lifetimes  $\tau_{\rm rest}<1 \; s$ to evade nucleosynthesis constraints \cite{bbn}.\\

As mentioned, inside the WD we assume that a fraction of DM is present. The population number of DM particles inside the star, $N_{\chi}$, will be the result of several processes. On one hand, the capture rate, $\Gamma_{\rm capt}$, collecting  particles by gravitational capture from an existing galactic DM distribution. It is indeed expected at early ages for the star  from the hydrogen dominated era, during the main sequence, and later, in the more compact configuration \cite{Garani,gould}.
 On the other hand, there exist other processes having the opposite effect, such as annihilation and evaporation, each with a  definite rate $\Gamma_{\rm ann}$, $\Gamma_{\rm evap}$, respectively, see for example \cite{zentner, kouv1, evap1, evap2}.
Let us emphasize that the strength of the possible (indirect) gamma ray  signal is to be directly related to the amount of DM that the star is able to capture and retain.

During the WD stage the DM capture rate can be  written \cite{Bramante, Kouvaris} as
\begin{equation}
\Gamma_{\rm capt}=\frac{\sqrt{24\pi}G\rho_\chi M_{WD}R_{WD}}{m_\chi \bar{v}}f_{\chi, A} \left[1-\frac{1-e^{-B^2}}{B^2} \right], 
\end{equation}
where $G$ is the gravitational constant, $M_{WD}$ and $R_{WD}$ are the WD mass and radius, $\rho_{\chi}$ is the local DM density, $m_{\chi}$ is the DM particle mass and $\bar{v}$ is the velocity dispersion between the DM particle and the WD. $f_{\chi, A}$ is the fraction of particles that undergo one or more scatterings with a nucleus of mass $m_A$ and baryonic number $A$ while inside the star. In our work we set $A\sim 14$ to account for a mixed C-O composition. $f_{\chi, A}$ saturates to unity  when it is larger than a typical scale $\sigma_{\rm sat}=\frac{\pi R_{WD}^2 m_A}{M_{WD}}$, known as geometrical cross section. Thus $f_{\chi, A} \sim 1$, if ${\sigma_{\chi, A}}\gtrsim {\sigma_{\rm sat}}$, where $\sigma_{\chi, A}\simeq A^2 \sigma_{\chi, N}$ is the DM-nucleus cross section \cite{lopes,hurst,amaro,cemb}, expressed in terms of the  DM-nucleon cross section, $\sigma_{\chi, N}$. Otherwise, $f_{\chi, A} \sim \frac{\sigma_{\chi, A}}{\sigma_{\rm sat}}$. Finally, the bracketed term accounts for DM that scatters but it is not captured in the WD, being 
\begin{equation}
B^2=\frac{6m_\chi v_{\rm esc}^2}{m_A \bar{v}^2 \left(\frac{m_\chi}{m_A}-1 \right)^2 },
\end{equation}
and 
\begin{equation}
v_{\rm esc}=\sqrt{\frac{2GM_{WD}}{R_{WD}}},
\end{equation}
the escape velocity.\\

Additional decay or even co-annihilation with a different species could be possible but, for the sake or brevity, we will not consider those channels here. We assume no DM self-interaction exists. In order to check the consistency of our argument neglecting  evaporation for light DM, we can estimate the limiting evaporation mass by demanding that the typical DM particle velocity, $v\sim \sqrt{\frac{T_c}{m_\chi}}$, at the WD central temperature, $T_c$, be less than the local escape velocity of the star, $v_{\rm esc}$. Thus, the evaporation mass value will somewhat depend on the thermodynamical properties of the star considered. For the WD masses $M_{WD} \in [0.2 M_\odot,0.95 M_\odot]$ and radii $R_{WD}\in [ 1.28 \,10^{-2}R_\odot,2.15 \,10^{-2} R_\odot]$ evaporation mass limits attain values $m_{\chi, \rm evap}\sim \frac{T_c}{v_{\rm esc}^2}\sim 2\; \rm MeV$ when we set  $T_c\sim 10^6 \; \rm K$. The corresponding set of WD masses and radii have been obtained using  the Lane-Emden solution for the non-relativistic polytropic equation of state with $n=3/2$ \cite{liu}. Inside WDs mass density can be reasonably taken as a slowly variating radial function, approximately equal to the central density $\rho(r)\sim \rho_c$. In our case the specific central density value range we use to obtain the selected $M_{WD}-R_{WD}$ parameter space is  $\rho_c \in[1.66 \, 10^5, 3.78 \, 10^6]\, \rm g/cm^3$.

As explained, in the range of $\chi$ masses we will consider, well above evaporation limits, the number of DM particles in the WD is obtained by solving the differential equation
\begin{equation}
\frac{dN_\chi}{dt}=\Gamma_{\rm capt}- 2\Gamma_{\rm ann},
\label{dN/dt}
\end{equation}
where 
\begin{equation}
\Gamma_{\rm ann}=\frac{1}{2}\int d^3\vec{r}\; n_\chi^2(\vec{r})\langle \sigma_a v\rangle=\frac{1}{2} C_a N_\chi^2,
\label{cap}
\end{equation}
and $\langle \sigma_a v\rangle$ is the annihilation cross section averaged over the initial DM states and $n_\chi(\vec{r})$ is the DM number density inside the star, which verifies $\int d^3\vec{r} \; n_\chi(\vec{r})=N_\chi$ at a given time. For simplicity the $\langle \sigma_a v\rangle$ will be taken in what follows as that of a wimp-like candidate $\langle \sigma_a v\rangle \sim 10^{-26}$ $\rm cm^3/s$ but this parameter will indeed depend on the interacting nature of the DM candidate and will discuss later on its impact in our calculation. In particular this value will assure that during the typical  WD lifetime of $\sim$ Gyr, DM can effectively thermalize in the interior of the star. In such a case the DM particle number density can be cast under a Gaussian form 
\begin{equation}
n_{\chi}(r)=n_{0,\,\chi} e^{-\left(\frac{r}{r_{\rm th}}\right)^2},
\end{equation} 
where $n_{0,\,\chi}$ is the central number particle density value and the thermal radius is given \cite{Kouvaris, PPPC4} by 
\begin{equation}
r_{\rm th}=\sqrt{\frac{9T}{8\pi G 
\rho_c m_\chi}}.
\end{equation}
Using this expression we can find an explicit form for $C_a\sim \langle \sigma_a v\rangle /r^3_{\rm th}$ in Eq.(\ref{cap}). Finally, solving Eq. \eqref{dN/dt}, it is found that $\Gamma_{\rm ann} \simeq \frac{\Gamma_{\rm capt}}{2}$ for times much larger than the typical  equilibrium time scale $\tau_{\rm eq}={1/\sqrt{\Gamma_{\rm capt}C_a}}$.  

\subsection{Dark mediator attenuation}

In this paper we are interested in the detectable signature of a mechanism of DM annihilation through dark mediators that subsequently decay to a photon pair, and its impact on the gamma ray luminosity and flux. 

In \cite{Leane} expressions for solar photon energy flux due to DM annihilation with free-streaming mediators were given
\begin{equation}
E_\gamma^2 \frac{d\Phi}{dE_\gamma}=\frac{\Gamma_{\rm ann}}{4 \pi d^2} E_\gamma^2 \frac{dN_\gamma}{dE_\gamma} P^Y_{\rm d, out},
\label{lea}
\end{equation}
where
$P^Y_{\rm d, out}$ is the probability that the mediator decays at a distance $d$ outside the WD. In what follows we explain for our different scenario how we include attenuation and decay effects, as they are not explicitly reflected in Eq.(\ref{lea}). The energy spectrum in the decay process $Y\rightarrow \gamma \gamma$  is given by a box-type shape
\begin{equation}
 \frac{dN_\gamma}{dE_\gamma}=\frac{4}{\Delta E} \Theta (E_\gamma-E_-) \Theta (E_+-E_\gamma) ,
\label{spec}
\end{equation}
as described in \cite{ibarra1, ibarra2}, with $\Delta E=E_+-E_-$ and 
\begin{equation}
E_\pm=\frac{1}{\gamma_Y}\frac{m_Y}{2}(1\mp v_Y)^{-1}.
\end{equation}
In this last expression the mediator velocity can be written as 
\begin{equation}
v_Y=\frac{p_Y}{E_Y},
\end{equation}
where $m_Y$, $p_Y$ and $E_Y$ are the mass, momentum modulus and energy of the mediator in the medium, respectively.  Note that each of the four photons emitted per annihilation has a monochromatic energy in the rest frame of the mediator,
\begin{equation}
E_{\gamma, \rm rest}=\frac{m_Y}{2}.
\end{equation}
\begin{figure}[tp]
\begin{center}
\includegraphics [angle=0,scale=1.3] {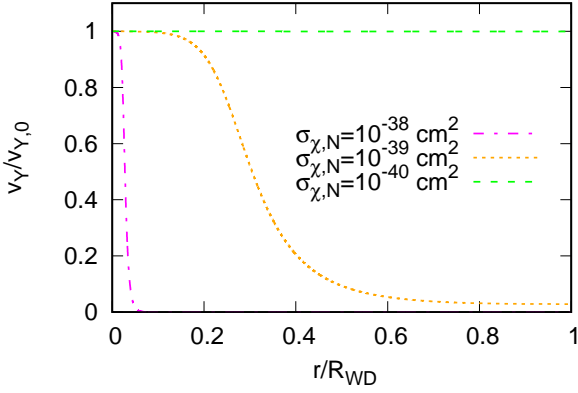}
\caption{In-medium dark mediator velocity as a function of the radial distance inside the WD for $m_\chi=500\;\rm MeV$ and $m_Y=10\;\rm MeV$. We fixed $\sigma_{Y, N}=\sigma_{\chi, N}=10^{-38}, 10^{-39}, 10^{-40}\; \rm cm^2$ with dot-dashed, dotted and dashed lines, respectively. See text for details.}
\label{fig11}
\end{center}
\end{figure}

Therefore, in the laboratory frame, where DM particles move non-relativistically, the photon energy can be written as 
\begin{equation}
E_\gamma=\frac{1}{\gamma_Y}E_{\gamma, \rm rest}(1-v_Y \cos\,\theta)^{-1},
\end{equation}
and, since the mediator decays isotropically, the resulting energy spectrum presents a box-shaped structure with a photon energy $E_\gamma \in[E_-,E_+]$ as obtained in Eq.(\ref{spec}).

In order to consider the fact that the mediator may suffer  energy attenuation when passing through the medium, we consider both momentum and energy will depend on the distance to the center of the star $r$, thus $p_Y\equiv p_Y(r)$ and $E_Y\equiv E_Y(r)$. It is important to remark here that for the stellar conditions under inspection $r_{\rm th}\ll R_{WD}$ and therefore we will approximate DM particles annihilate at $r\sim 0$ .

Under these circumstances when the mediator particle is created in the boosted scenario we have, initially, the momentum modulus $p_{0,Y}=\sqrt{m_\chi^2-m_Y^2}$. Later, when the mediator interacts inside the WD it suffers a number of elastic scatterings that could be approximated as a continuous energy-loss with a small-deflection angle \cite{Ritz,starkman,kouv3}. It is therefore reasonable to write the fractional momentum loss in terms of the variation of the mediator momentum prior to the interaction using a coefficient, $q\sim (m_A-m_Y)/m_A$ with $0<q<1$ parameterizing the collision so that when the $Y$ particle has traveled a distance $r$ 
\begin{equation}
\frac{dp_Y}{dr}=\frac{\Delta p_Y}{\lambda_{I}}=\frac{-(1-q)p_Y}{\lambda_{I}},
\label{dpdr}
\end{equation} 
where $\lambda_{\rm I}=\frac{1}{\sigma_{Y,A}n(r)}$,  $\sigma_{Y,A}$ is the nucleus-mediator cross section. Assuming that the main contribution to the cross section of nuclei comes from the coherent enhancement of the spin-independent cross section we can further consider  $\sigma_{Y,A}=A^2 \sigma_{Y, N}$, being $\sigma_{Y, N}$ the nucleon-mediator cross section. At this point we must note that while the DM-nucleon cross section is much  constrained from current direct searches, it is scarcely tested for  mediators. We will assume in what follows that due to the secluded nature of these type of DM mediators they will couple to nucleons with less (or up to  the same) strength as compared to DM particles $\sigma_{Y,N}\sim (10^{-3}-1) \sigma_{\chi, N}$.

We denote $n(r)=\frac{\rho_c}{A m_N}\int_0^r \omega(r')^{\frac{3}{2}} dr'$ the number density of nuclei in the stellar profile. Since the supporting pressure in the WD is provided by the degenerate electron fraction to obtain $n(r)$ we use a polytropic approach to the equation of state and approximate the analytic solution of the Lane-Emden equation with a polytropic index $n=\frac{3}{2}$ following \cite{liu} as
\begin{eqnarray}
\omega(r)&=&-\alpha(1+B_{3/2}\xi^2)^{-2}+(1+\alpha)\left( 1+ \frac{1}{12}\xi^2\right)^{-2} \nonumber \\  & + & \frac{\alpha}{6}\xi^2 \left(1+\frac{1}{12}\xi^2 \right)^{-3}+\frac{4.6\; 10^{-4}\xi^3}{(1+1.1\; 10^{-3}\xi^2)^2},
\end{eqnarray}
with $\xi=\frac{r}{a}$, $a^2=\frac{5K}{8\pi G}\rho_c^{\frac{-1}{3}}$ and $K=P_c/\rho_c^{\frac{5}{3}}$, given the central density and pressure values $\rho_c$, $P_c=\frac{(3 \pi^2)^\frac{2}{3}}{5m_e}\left[ \left(\frac{Z}{A} \frac{\rho_c}{m_N} \right)  \right]^\frac{5}{3}$, respectively. $m_e, m_N$ are the electron and nucleon masses. 
Using the prescribed fit with $\alpha=0.481$, $B_{3/2}=\frac{18}{5}\left( \frac{4\alpha}{4+5\alpha}\right)^4$ one can obtain a convenient  approximation for the the full numerical solution with an accuracy of $1\%$.
Integrating Eq.(\ref{dpdr}) we obtain
\begin{equation}
p_Y(r)=\sqrt{m_\chi^2-m_Y^2}\; e^{\frac{-(1-q)A\sigma_{Y,N}\rho_c}{m_N}\int_0^r \omega(r')^{\frac{3}{2}}dr'},
\end{equation}
and accordingly, the radial dependent energy is given by $E_Y(r)=\sqrt{p_Y(r)^2+m_Y^2}$. From this expression it is clear that the energy spectrum, $E_\gamma^2 \frac{dN_\gamma}{dE_\gamma}$, will be attenuated with radial distance from the production site inside the star.

In Fig.(\ref{fig11}) we show the dark mediator  attenuation from initial velocity $v_{Y,0}=\frac{\sqrt{m_\chi^2-m_Y^2}}{m_\chi}$ as a function of the distance inside the WD for $m_\chi=500\;\rm MeV$ and $m_Y=10\;\rm MeV$. We consider  $\sigma_{Y, N}=\sigma_{\chi, N}=10^{-38}, 10^{-39}, 10^{-40}\; \rm cm^2$ depicted with  dot-dashed, dotted and dashed lines, respectively. The WD configuration corresponds to $\rho_c=3.776 \times 10^6$ $\rm g/cm^3$ and $R_{WD}=1.28\, 10^{-2} R_{\odot}$, $M_{WD}=0.95M_{\odot}$. In order to emphasize the effect we have imposed $\sigma_{Y, N}=\sigma_{\chi, N}$ but we will discuss later this dependence.

\subsection{Photon luminosity from dark decay}

The equation which governs the decay probability density of the mediator inside the star can be written as
\begin{equation}
\frac{dP_{\rm dec}}{dr}=\frac{-P_{\rm dec}}{\gamma_Y \tau_{\rm rest}}=\frac{-P_{\rm dec}m_Y}{\tau_{\rm rest} E_Y(r)},
\end{equation}
where the relativistic decay length $\tau=\gamma_Y\tau_{\rm rest}$.

The decay probability must fulfill the normalization condition $\int_0^\infty P_{\rm dec}dr=1$. Explicitly,
\begin{equation}
N \int_{0}^\infty e^{-\int_0^r \frac{m_Ydr'}{\tau_{\rm rest}E_Y(r')}}dr=1,
\label{N}
\end{equation}
where $N$ is a normalization constant. Beyond the stellar radius we will be assuming no energy losses such that the mediator will not be further attenuated and its energy  remains quenched $E_Y(r)=E_Y(R_{\rm WD})$, $r>R_{\rm WD}$. If further scattering with an external agent was introduced  an additional attenuation would arise \cite{Ingelman}. For the sake of clarity we will not consider that refinement here.

From integration in Eq.(\ref{N}) we can obtain the actual form for the probability $P_{\rm dec,d>R_{WD}}\equiv P^Y_{\rm d,out}$that the mediator decays outside the star, between $R_{\rm WD}$ and a generic distance $d>R_{\rm WD}$
\begin{equation}
P_{\rm dec,d>R_{WD}}=\int_{R_{\rm WD}}^d N \; e^{-\int_0^r \frac{m_Y dr'}{\tau_{\rm rest}E_Y(R_{\rm WD})}}dr,
\end{equation}
or explicitly,
\begin{equation}
P^Y_{\rm d,out}={\mathcal P} 
\left( 1- e^{-\frac{m_Y(d-R_{\rm WD})}{\tau_{\rm rest} E_Y(R_{\rm WD})}}\right), 
 \label{Pout}
\end{equation}
where we have used 
\begin{equation}
{\mathcal P}=\frac{N \tau_{\rm rest}E_Y(R_{\rm WD})}{m_Y} e^{-\frac{m_YR_{\rm WD}}{\tau_{\rm rest} E_Y(R_{\rm WD})}}. \end{equation}
\begin{figure}[tp]
\begin{center}
\includegraphics [angle=0,scale=1.3] {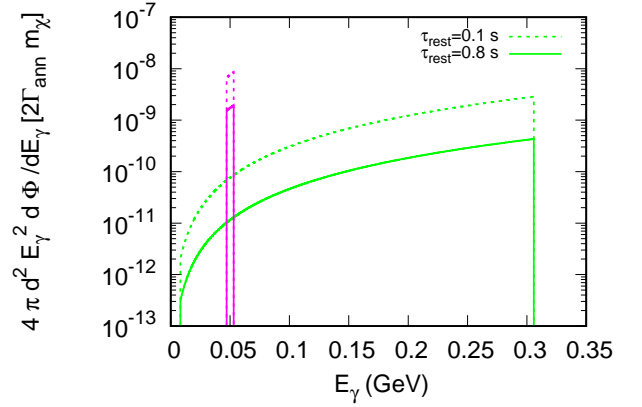}
\caption{Energy flux at distance $d=2R$ due to DM annihilation as a function of the photon energy for different values of $\tau_{\rm rest}=0.8$ s (solid lines) and $\tau_{\rm rest}=0.1$ s (dashed lines). It is normalized to the product $2 \Gamma_{\rm ann} m_\chi$. We consider two values for the $\sigma_{\chi, N}=10^{-39}$ $\rm cm^2$ (green) and $\sigma_{\chi, N}=5\times 10^{-39}$ $\rm cm^2$ (magenta). We assume $\sigma_{\chi, N}=\sigma_{Y, N}$ and  fix $m_\chi=800$ MeV and $m_Y=100$ MeV. See text for details.}
\label{fig2}
\end{center}
\end{figure}
On the other hand, the probability of decay $Y \rightarrow \gamma \gamma$ inside the WD $(r<R_{\rm WD})$ weighted with the position dependent spectrum in Eq.(\ref{spec}) will allow to obtain the photon luminosities deposited inside the stellar volume and extract valuable information of the strength of attenuation in the stellar medium.

The internal luminosity $L_\chi$ due to annihilation of DM particles into photons through dark mediators inside the stellar volume can be thus written as
\small
\begin{equation}
L_\chi=\Gamma_{\rm ann} \int_0^{R_{\rm WD}} N \; e^{-\int_0^r \frac{m_Y dr'}{\tau_{rest}E_Y(r')}} \left[\int_{E_-(r)}^{E_+(r)} E_\gamma \frac{dN_\gamma(r)}{dE_\gamma}dE_\gamma \right] dr,
\end{equation}  
\normalsize
where now dependencies are made explicit. On one hand, the dependence on the radial coordinate of the energy spectrum of photons produced inside the WD as a result of the finite lifetime of the mediators and, on the other hand, the medium interaction of the mediators from the spatially dependent limiting values in the energy interval, $E_{\pm}(r)$. When $\sigma_{Y,N}\rightarrow 0$ we recover the case $p_Y(r)=p_{Y,0}$ where no attenuation is present.

\section{Results}

In this section we analyze the results obtained for the photon luminosity  and flux to compare with existing experimental measurements of the coldest WDs. By comparing the expected internal warming due to the dark mediator in-medium decay one could obtain some constraints to the lifetime of the mediator in the scenario considered.
In order to maximize the possible DM effects in the stellar warming we consider those WDs present in the M4 GC \cite{Bedin} where, in line with \cite{fairbairn}, we assume a DM density $\rho_\chi=2660\rho_{\chi,0}$, being $\rho_{\chi, 0}=0.3\; \rm GeV cm^{-3}$ the solar-circle value, usually taken as reference in our galactic DM distribution. The quoted value in M4 GC refers to the density at the largest radius where the WD data are observed \cite{spergel, fairbairn}, having a velocity dispersion of $\bar{v}=20\; \rm km/s$. It is important to mention that in this GC the age of the WDs is set to $t=12.7 \pm 0.7$ Gyr \cite{hansel}, so that the assumption that the DM particles which have some impact on the luminosity through annihilation are those which are captured in the WD stage seems a reasonable  hypothesis. In the same line, it is clear that at this age they are thermalized, $t>\tau_{\rm eq}$.

\begin{figure}[tp]
\begin{center}
\includegraphics [angle=0,scale=1.3] {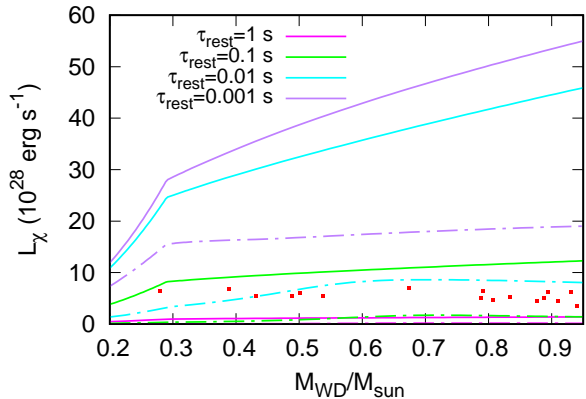}
\caption{WD internal luminosity due to DM annihilation as a function of the WD mass for different values of $\tau_{rest}=0.001,0.01,0.1,1$ s using $m_\chi=500 \; \rm MeV$ and  mediator masses, $m_Y=375\;\rm MeV$ (solid lines) and $m_Y=10 \; \rm MeV$ (dash-dotted lines). We fixed $\sigma_{\chi, N}=\sigma_{Y, N}=10^{-39}\; \rm cm^2$. Red points are experimental data for M4 GC from \cite{fairbairn}.}
\label{fig3}
\end{center}
\end{figure}

In Fig.(\ref{fig2}) we show the energy flux due to annihilation of DM with $m_\chi=800$ MeV and $m_Y=100$ MeV as a function of the photon energy at distance $d=2R$ for a WD with mass $M_{\rm WD}=0.28M_\odot$ and radius $R_{WD}=0.04R_\odot$ and normalized to the product $2 \Gamma_{\rm ann} m_\chi$. We consider two different values of $\tau_{\rm rest}=0.8$ s (solid lines) and $\tau_{\rm rest}=0.1$ s (dashed lines) and $\sigma_{\chi, N}=10^{-39}$ $\rm cm^2$ (green) and $\sigma_{\chi, N}=5\times 10^{-39}$ $\rm cm^2$ (magenta). We assume $\sigma_{\chi, N}=\sigma_{Y, N}$ and $q=0.5$. The case where more attenuation is obtained is that of the larger cross-section where the available energy window for photoproduction is small and centered about $E_\gamma=m_Y/2$. For the considered cases it is fullfilled that $\lambda_D>2R$ and as $\tau_{\rm rest}$ grows the flux obtained is smaller since less mediators have been able to decay into photons at the radial distance $2R$. 
\begin{figure}[t]
\begin{center}
\includegraphics [angle=0,scale=1.3] {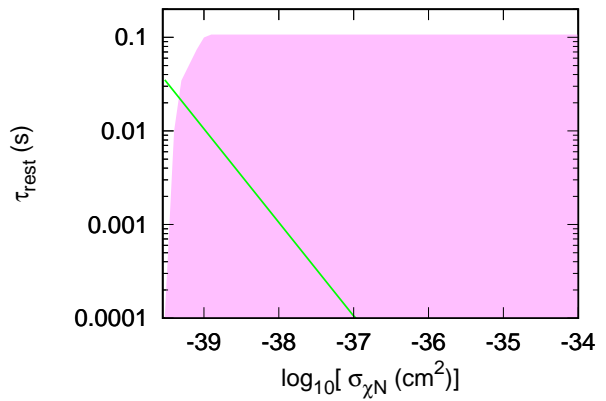}
\caption{Constraints for the mediator lifetime as a function of DM-nucleon cross section for the case $m_Y/m_\chi\rightarrow 1$. We fixed $\sigma_{\chi,N}=\sigma_{Y, N}$. Coloured region depicts the excluded parameter space, see the text for more details. Green line is a lower limit signaling where the mediator scatters at least once in the WD. }
\label{fig5}
\end{center}
\end{figure}

In Fig.(\ref{fig3}) we show the WD internal luminosity i.e. energy per unit time deposited inside the stellar volume, due to the annihilation of  DM as a function of the WD mass for $m_\chi=500 \; \rm MeV$ and $m_Y=375\;\rm MeV$ (solid lines) and $m_Y=10 \; \rm MeV$ (dash-dotted lines). We set  different lifetimes $\tau_{\rm rest}=0.001,0.01,0.1,1$ s. We fix $\sigma_{\chi, N}=\sigma_{Y, N}=10^{-39}\; \rm cm^2$ and $q=0.5$. We can see that the more similar to the DM particle mass, $m_\chi$, the mediator mass, $m_Y$, is, the higher the luminosity is. This happens because the mediator is produced almost at rest and it is equivalent to a prompt photoproduction (there is no energy attenuation for the mediator). Moreover, the luminosity decreases when $\tau_{\rm rest}$ increases due to the fact that the higher $\tau_{\rm rest}$ is, the smaller the probability of decaying inside the object. It is clear from the figure that we can exclude sets of parameters that yield an internal luminosity $L_\chi$ higher than the experimental extracted for WDs in M4 GC (red points as given by \cite{fairbairn}). In other words, for fixed values of $m_\chi$, $m_Y$ and $\sigma_{\chi, N}$ (that is fixed to be equal to $\sigma_{Y, N}$) there will be a limiting lower value of $\tau_{\rm rest}$ below which the luminosity would be higher than that deduced from experimental data. 

In order to analyse these specific constraints we plot in Fig.(\ref{fig5}) the excluded values of $\tau_{\rm rest}$ (coloured region) as a function of the logarithm (base 10) of $\sigma_{\chi, N}$ for $m_Y/m_\chi\rightarrow 1$. Being conservative, we exclude values of $\tau_{\rm rest}$ which provide luminosities beyond a $50\%$ tolerance for the complete set of all  experimental data, i.e. $L_\chi> 1.5 \; L_{\rm exp}$. We fix this error band since experimentally deduced luminosities for WDs are accurate only to the first or second significant figures. Incidentally, this happens for luminosities above $50\%$ of the value for the first experimental data point considered in the series ($M_{WD}=0.28M_\odot$, $R_{WD}=2.7\times 10^9$ cm for $\rho \sim 3.3 \times 10^{5}$ $\rm g/cm^3$). On this plot we show the boundary of the coloured region, whose physical meaning is that of the minimum value of $\tau_{\rm rest}$  below  which lifetimes for a decaying mediator produced at rest are not allowed for a given $\sigma_{\chi, N}$. In this case since the mediator decays at rest does not suffer attenuation so that there is no dependence on $\sigma_{Y, N}$. As it is obtained it is indeed a lower limit of allowed $\tau_{\rm rest}$ independent on $m_\chi$, $m_Y$ and $\sigma_{Y, N}$. Besides, the green line indicates the corresponding Y-lifetime where $\tau_{\rm rest}=\lambda_{Y,A}$, i.e. the value for which at least one scattering between the mediator $Y$ and a nucleus $A$ inside the WD will take place and assuming $\sigma_{Y,N}=\sigma_{\chi, N}$. Below this limiting value situation is equivalent to that with no energy losses.
In the figure it is shown that for saturated values of the capture rate in the WD i.e. $\sigma_{\chi, N}>\sigma_{\rm sat}=1.1\,10^{-39}\; \rm cm^2$ there is no further change in the limiting value of  $\tau_{\rm rest}$. We refer to this as $\tau_{\rm limit}$. For smaller values $\sigma_{\chi, N}<\sigma_{\rm sat}$ there is a quenching of $\tau_{\rm limit}$ as smaller values of $\tau_{\rm rest}$ are not excluded from luminosity constraints. We expect that our results could be in principle extended up to $\sigma_{\chi, N}\sim 10^{-29}\; \rm cm^2$ covering the targeted region in the phase space for sub-GeV DM as we comment in the Introduction section.
\begin{figure}[tp]
\begin{center}
\includegraphics [angle=0,scale=1.3] {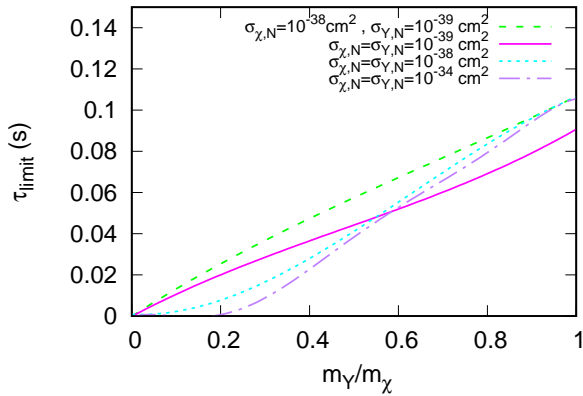}
\caption{$\tau_{\rm limit}$ (boundary of excluded $\tau_{\rm rest}$ region) as a function of the ratio  $m_Y/m_\chi$ for different values of $\sigma_{\chi,N}$,$\sigma_{Y, N}$. See text for details.}
\label{fig4}
\end{center}
\end{figure}
In other more general cases, for a fixed value of $\sigma_{\chi, N}$ and given $m_\chi, m_Y$ the maximum excluded value of $\tau_{rest}$, $\tau_{\rm limit}$, will be lower than that obtained in the extreme $m_Y \sim m_\chi$ case. In other words, the most restrictive $\tau_{\rm rest}$ limit would be that in which is independent on $\sigma_{Y,N}$ and $m_\chi, m_Y$.

In Fig.(\ref{fig4}) $\tau_{\rm limit}$ (boundary of the excluded region for $\tau_{\rm rest}$) is shown as a function of the ratio  $m_Y/m_\chi$ for different values of $\sigma_{\chi,N}$, $\sigma_{Y, N}$. In these cases when $m_Y/m_\chi\neq 1$ there is effective contribution from energy attenuation and decay effects. We fixed $\sigma_{\chi, N}=\sigma_{Y, N}=10^{-39}\; \rm cm^2$ (solid lines), $\sigma_{\chi, N}=\sigma_{Y, N}=10^{-38}\; \rm cm^2$ (short-dashed lines), $\sigma_{\chi, N}=\sigma_{Y, N}=10^{-34}\; \rm cm^2$ (dash-dotted lines) and $\sigma_{\chi, N}=10 \sigma_{Y, N}=10^{-38}\; \rm cm^2$ (long-dashed  lines). We can see that the smaller the ratio $m_Y/m_\chi$ is, the smaller the lifetime limit is,  since the photons would yield luminosities compatible with experimental bounds, thus being a weak constraint. In the case of $\sigma_{\chi, N}\lesssim 1.1 \; 10^{-39}\; \rm cm^2$ (solid line) there is a further effect not present in the other cases considered due to the reduction of the WD capture rate of DM as the saturation factor $f_{\chi,A} <1$. It can be seen that the effect of reducing $\sigma_{Y, N}$ fixing $\sigma_{\chi, N}$ is quenching $\tau_{\rm limit}$.

\section{Conclusions}

We have studied the gamma ray emission in WDs from annihilation of DM in their interior through metastable mediators. We have considered the combined effect of energy attenuation and finite decaying lifetime. Using an approximation where the energy loss can be described in a continuous way through mediator scattering with nuclei inside the stellar volume we have derived the internal luminosities and fluxes. We have compared these luminosities to those from cold WDs in M4 GC. We find that in the case where $\sigma_{\chi, N}\lesssim 10^{-40}\; \rm cm^2$  the attenuation is negligible and the only effect comes determined from the lifetime. However for larger cross-sections, up to $\sigma_{\chi, N}\sim 10^{-29}\; \rm cm^2$ there are non-trivial effects that further constrain the lifetime bounds with a monotonic increase in the $m_Y/m_\chi$ ratio. The effect of the expected seclusion of DM from nucleons can be seen by imposing the non-degeneracy of  $\sigma_{\chi, N}$ and $\sigma_{Y, N}$ values. We find that the smaller the $\sigma_{Y, N}$ is the less restrictive effect on the allowed Y-lifetimes. Note that the annihilation rate of DM or, in other words,  the value of the thermally averaged annihilation cross section could somewhat modify our results concerning obtained luminosities and flux (and lifetimes) as an increase factor would yield larger values of $\tau_{\rm limit}$. As for energy flux we find the more attenuation there is the sharpest photoproduction results. 
Let us keep in mind that the considerations in the discussion about the restrictions on the decaying lifetimes for mediators from cold WDs must take into also account, apart from other astrophysical scenarios, the current efforts in colliders. As a new strategy to follow they have shifted towards an alternative simplified model paradigm that includes these additional mediators and the search for a displaced secondary vertex, characterised by the mass of the particle and its lifetime, see \cite{Buchmu}. This has led to an extensive effort amongst both theorists and experimentalists at the LHC to establish a systematic programme to characterise DM searches using simplified models. A multi-directional and complementary approach from different search contexts will most surely provide valuable information on this type of models. 

\section*{Acknowledgments}
We thank useful discussions with R. Lineros, M. Ardid, M. A. S\'anchez-Conde and J. Silk. This work has been supported by Junta de Castilla y Le\'on SA083P17, FIS2015-65140-P projects and by PHAROS Cost action. We also thank the support of the Spanish Red Consolider MultiDark FPA2017‐90566‐REDC M. Cerme\~no is supported by a fellowship from the University of Salamanca.

\end{document}